\documentstyle[twoside,fleqn,espcrc2]{article}
 
\input psfig
 
\newcommand{\OO}{{\cal O}}
\newcommand{\BB}{{\cal B}}
\newcommand{\bBB}{\bar{\cal B}}

\newcommand{\AmS}{{\protect\the\textfont2
  A\kern-.1667em\lower.5ex\hbox{M}\kern-.125emS}}

\title{A Test of Target Independence of the `Proton Spin' Effect in
Semi-Inclusive Deep Inelastic Scattering}

\author{G.M. Shore \address{Department of Physics,
University of Wales Swansea, \\
Singleton Park, Swansea SA2 8PP, U.K. }
\thanks{SWAT-96/130}
\thanks{Invited talk at QCD 96, Montpellier, July 1996}  }
 
\begin{document}
 
\begin{abstract}
The target-independent suppression mechanism proposed as an 
explanation of the EMC-SMC `proton spin' effect is reviewed and 
compared to current experimental data. According to this proposal, 
the anomalous suppression 
observed in the first moment of the polarised proton structure 
function $g_1^p$ is not a special property of the proton structure, 
but reflects an anomalously small value of the first moment of the
QCD topological susceptibility. It is shown how this mechanism 
can be tested in semi-inclusive DIS processes in which a pion or $D$
meson carrying a large target energy fraction $z$ is detected
in the target fragmentation region.
\end{abstract}
 
\maketitle
 
\section{Introduction}
The so-called `proton spin' problem, viz.~the anomalous suppression
in the first moment  $\Gamma_1^p$ of the polarised structure function $g_1^p$,
continues to be the focus of intense theoretical and experimental activity.
In the explanation proposed in collaboration with S.~Narison and
G.~Veneziano\cite{SV,NSV}, this suppression is identified as a generic, 
target-independent feature of QCD related to the axial anomaly and not
a special property of the proton structure.\footnote{It is also emphasised
in \cite{SV,NSV} and elsewhere (e.g.~\cite{AR}), 
yet consistently ignored by the experimental community, 
that the singlet form factor $\Delta \Sigma$ does {\it not} measure spin.}  
The violation of the Ellis-Jaffe sum rule for the flavour singlet form factor 
$G_A^{(0)}$ ($\Delta \Sigma$ in standard notation) 
contributing to $\Gamma_1^p$ is understood as one more example of 
the class of OZI-violating phenomena characteristic of the flavour-singlet 
pseudoscalar or pseudovector channels. More specifically, it is shown to reflect
an anomalous suppression in the first moment of the QCD topological
susceptibility $\chi^{\prime}(0)$.

In our view, this explanation is well-motivated and theoretically convincing.
However, in view of the difficulties, both theoretical and especially experimental,
in obtaining sufficiently accurate numbers for $\Gamma_1^p$ to give a
conclusive confirmation of our proposed mechanism, it is interesting
to look for other indirect tests. One characteristic, though not 
unique, feature of our analysis
is the prediction of target independence. In this talk, after briefly reviewing
our earlier work and the current experimental status of the $\Gamma_1^p$
sum rule, I shall describe some ideas\cite{SV2} on how the hypothesis of 
target-independent flavour-singlet suppression can be tested experimentally
in semi-inclusive deep inelastic scattering in which a pion or $D$ meson
carrying a large target energy fraction $z$ is detected in the target
fragmentation region. 
 
\section{The sum rule for $\Gamma_1^p$}
 
The sum rule for the first moment
of the polarised proton structure function $g_1^p$ is\footnote{We work 
throughout in the chiral limit. Since all our results depend smoothly on the light 
quark mass terms in this limit\cite{SV,NSV}, the corrections induced by finite quark
masses will be small.}:
\begin{eqnarray}
\Gamma^p_1(Q^2) &&\equiv
\int_0^1 dx g_1^p(x;Q^2) \cr
&&= {1\over 6}  \biggl[ \biggl(G_A^{(3)}(0)
+ {1\over \sqrt3}G_A^{(8)}(0)\biggr)
R(\alpha_s)\cr
&&~~~~~~~~~~~~+ {2\over3} G_A^{(0)}(0;Q^2)
R_0(\alpha_s) \biggr]
\end{eqnarray}
where $G_A^{(a)}(k^2)$ are the form factors in the proton
matrix elements of the axial current:
\begin{equation}
\langle P|J^a_{\mu 5R}(k)|P\rangle =
G_A^{(a)} \bar u \gamma_\mu \gamma_5 u +
G_P^{(a)} k_\mu \bar u \gamma_5 u
\end{equation}
and $R(\alpha_s)$ and $R_0(\alpha_s)$ are known perturbative series.
The suffix `$R$' indicates a renormalised composite operator
and $a$ is an $SU(3)$ flavour index. 
Since the flavour singlet current $J_{\mu 5R}^0$
is multiplicatively renormalised, its matrix elements, in particular
$G_A^{(0)}(0;Q^2)$, are renormalisation group (RG) scale dependent, 
scaling with anomalous dimension $\gamma$,
a fact which plays a crucial r\^ole in the correct interpretation of 
the `proton spin' effect. 

In the QCD-improved parton model, these form factors are related
to the polarised quark and gluon distributions by (see, e.g.~\cite{AR})
\begin{eqnarray}
&&G^{(3)}_A(0) = {1\over2}(\Delta u-\Delta d) \cr
&&G^{(8)}_A(0) = {1\over{2\sqrt 3}}(\Delta u+\Delta d-2\Delta s)  \cr
&&G^{(0)}_A(0;Q^2) \equiv \Delta\Sigma = \Delta q^S - N_F {\alpha_s
\over 2\pi} \Delta g
\end{eqnarray}
In this latter expression, $\Delta q^S = \Delta u+\Delta d+\Delta s$ is assumed
to be RG invariant, while the polarised gluon distribution $\Delta g$ has
the complicated RG behaviour necessary to ensure multiplicative
scaling of $G_A^{(0)}(0;Q^2)$ ~(c.f.~\cite{SV}).   

The Ellis-Jaffe sum rule is simply the OZI prediction,
\begin{equation}
G_A^{(0)}(0)\big|_{\rm OZI} = 2\sqrt3 G_A^{(8)}(0) = 0.58\pm0.02
\end{equation}
which leads to
\begin{equation}
\Gamma_1^p|_{Q^2=10GeV^2} = 0.170 \pm 0.003
\end{equation}
It is obtained in the parton model by neglecting the gluon distribution
$\Delta g$ in the singlet form factor and arguing that the polarised strange quark
distribution in the proton vanishes, i.e.~$\Delta s = 0$.

This sum rule is clearly violated by the experimental data. The most recent 
published results from the SMC collaboration on the proton structure 
function\cite{SMC} are 
\begin{equation}
\Gamma_1^p|_{Q^2=10GeV^2} = 0.136 \pm 0.011(\rm stat) \pm
0.011(\rm syst)~~~
\end{equation} 
from which we extract
\begin{equation}
G_A^{(0)}(0)\big|_{Q^2=10GeV^2} = 0.29 \pm 0.15
\end{equation}
However, a recent analysis by Ball, Forte and Ridolfi\cite{BFR}, 
taking into account the $Q^2$ evolution in extracting $\Gamma_1^p$ from 
the data (see also Deshpande in these proceedings)
and making a different extrapolation to the unmeasured small $x$ region,
leads to a lower central value, $\Gamma_1^p = 0.122 \pm 
0.013(\rm exp) +0.011/-0.005(th)$, while increasing the error estimate. 
Interestingly, they also find support in the data for a decomposition of 
$G_A^{(0)}$ into quark and gluon distributions in which the RG invariant 
$\Delta q^S$ is close to the OZI value, in accord with our expectation (see 
section 3 and \cite{SV,NSV}) that OZI violations probably reside in the 
non-RG invariant quantities.

Our prediction,
incorporating a QCD spectral sum rule estimate of the first moment of the
topological susceptibility, is 
\begin{equation}
\Gamma_1^p(Q^2=10GeV^2) = 0.143 \pm 0.005
\end{equation}
obtained from
\begin{equation}
G_A^{(0)}(0)\big|_{Q^2=10GeV^2} = 0.35 \pm 0.05
\end{equation}

Despite the large experimental errors, our prediction is clearly in good
agreement with the quoted SMC result, although the BFR\cite{BFR} reanalysis
of the data would move $\Gamma_1^p$ uncomfortably low. 
New data is expected shortly from the SMC collaboration and it will be 
interesting to see how this affects the comparision of theory with experiment.
A conservative assessment at present  is that the data unquestionably show an 
OZI violation in a channel where it is to be expected and of roughly the
magnitude that can be inferred, as in \cite{NSV}, from the known OZI breaking
in the singlet pseudoscalar channel reflected in the $\eta^{\prime}$ mass.
Viewed in this light, there is no `proton spin' {\it problem}, simply a
challenge to derive a quantitative prediction from first principles of a subtle, 
anomaly-related QCD phenomenon.

\section{Target Independence and the Composite Operator/Proper 
Vertex Method for DIS}

The field-theoretic methods we have used to derive (8), (9) are fully
described in \cite{SV,NSV} and summarised in \cite{S}. For a generic structure
function sum rule, we have the expression
\begin{eqnarray}
&&\int_0^1 dx  x^{n-1} F(x;Q^2)  \cr
&&= \sum_i \sum_j C_i^n(Q^2)~
\langle 0|\OO_i(0) \tilde\OO_j(0) |0\rangle~
\Gamma_{\tilde\OO_j N\bar N}~~~~~~
\end{eqnarray}
Here, $C_i^n$ are the Wilson coefficients and $\OO_i^n$ the lowest twist, 
spin $n$ operators appearing in the OPE of two electromagnetic currents.
The novel feature is our decomposition of the resulting nucleon matrix
elements $\langle N|\OO_i^n|N\rangle$ into the product of  zero-momentum
composite operator propagators $\langle 0|\OO_i(0) \tilde\OO_j(0) |0\rangle$
and  proper vertices $\Gamma_{\tilde\OO_j N\bar N}$ defined as 1PI with respect
to an appropriately chosen set of composite operators $\tilde\OO_j$. It is important to recognise that this decomposition is exact and no 
approximation is involved in using a finite (in practice small) set of operators. 
A different choice of the set $\tilde\OO_j$ merely changes the definition of the
vertices $\Gamma_{\tilde\OO_j N N}$ -- the trick is to choose a basis
of operators such that the resulting proper vertices have simple RG properties 
and ideally are closely related to physical couplings.

In the case of the $\Gamma_1^p$ sum rule, the relevant proton matrix 
element (after using the axial anomaly and assuming the absence of a
massless pseudoscalar Goldstone boson in the singlet channel) is simply the
forward matrix element of the gluon topological density
$\langle P|Q_R(0)|P\rangle$. In fact,
\begin{eqnarray}
G_A^{(0)}(0;Q^2) \bar u \gamma_5 u =
{1\over2M} 2N_F \langle P|Q_R(0)|P\rangle
\end{eqnarray}
(Here, $Q_R$ is the renormalised operator related by a non-multiplicative
renormalisation to the bare operator 
$Q_B = {\alpha_s\over8\pi} {\rm tr} G^{\mu\nu}\tilde G_{\mu\nu}$.)  
The composite operator basis 
$\tilde\OO_j$ is then chosen to be the set of renormalised flavour singlet
pseudoscalar operators $Q_R$ and $\Phi_{5R}$, where (up to a
subtle normalisation factor) the corresponding bare operator
$\Phi_{5B}$ is $i\sum\bar q \gamma_5 q$. We then have
\begin{eqnarray}
&&\langle P|Q_R(0)|P\rangle  \cr
&&= \langle 0|Q_R Q_R|0\rangle \Gamma_{Q_R P \bar P} +
\langle 0|Q_R \Phi_{5R}|0\rangle \Gamma_{\Phi_{5R} P \bar P}~~~~~~
\end{eqnarray}
where the composite operator propagators are at zero momentum
and the proper vertices are 1PI with respect to $Q_R$
and $\Phi_{5R}$ only.

The composite operator propagator in the first term is the zero-momentum
limit of the QCD topological susceptibility $\chi(k^2)$, viz.
\begin{equation}
\chi(k^2) = \int dx e^{ik.x} i\langle 0|T^* Q_R(x)~Q_R(0)|0\rangle
\end{equation}
The anomalous chiral Ward identities show that 
$\chi(0)$ vanishes for QCD with massless quarks,
in contrast to pure Yang-Mills theory where
$\chi(0)$ is non-zero. Furthermore, it can be shown\cite{SV} that the  
propagator $\langle 0|Q_R~\Phi_{5R}|0\rangle$ at zero momentum
is simply the square root of the first moment of the topological
susceptibility. We therefore find:
\begin{equation}
\langle P| Q_R(0) |P\rangle = 
\sqrt{\chi^{\prime}(0)} ~\Gamma_{\Phi_{5R} P \bar P}
\end{equation}
The quantity $\sqrt{\chi^\prime(0)}$ is
not RG invariant and scales with the anomalous dimension $\gamma$.
On the other hand, the proper vertex has been chosen specifically
so as to be RG invariant. The renormalisation group properties of this
decomposition are crucial to our resolution of the `proton spin problem'.
 
Our proposal (which is fully motivated in \cite{SV,NSV,S})
is that we should expect the source of OZI violations
to lie in the RG non-invariant, and therefore anomaly-sensitive, 
terms, i.e. in $\chi^{\prime}(0)$.
Our quantitative prediction (9) then follows by using the OZI approximation
for the vertex  $\Gamma_{\Phi_{5R} P \bar P}$ and a QCD spectral sum 
rule estimate of the first moment of the topological susceptibility.
We find\cite{NSV}  
\begin{equation}
\sqrt{\chi^\prime(0)}\Big|_{Q^2=10 GeV^2}
= 23.2 \pm 2.4~{\rm MeV}~,
\end{equation}
a suppression of approx.~$0.6$ relative to the OZI value
$f_\pi /\sqrt6$. 

Since, according to this proposal, the anomalous suppression in $\Gamma_1^p$
is assigned to the composite operator propagator rather than the
proper vertex, the suppression is a {\it target independent} property of
QCD related to the axial anomaly, {\it not} a special property of the
proton structure.  This immediately raises the question whether it is
possible to test our proposed mechanism by effectively performing DIS 
experiments on other hadronic targets.

\section{Other Targets}

Leaving until section 5 the question of how this may be experimentally
realised, we now consider the implications of this 
target-independent suppression mechanism for structure functions of
other hadrons besides the proton and neutron.

The basic assumption is that for any hadron, the singlet form factor in (1)
can be substituted by its OZI value times a universal (target-independent),
RG invariant, suppression factor $s(Q^2)$, i.e.
\begin{equation}
\Gamma_1 = {1\over6} R(\alpha_s)\biggl[ G_A^{(3)}(0) + {1\over\sqrt3}
G_A^{(8)}(0)\bigl(1 + 4s\bigr)\biggr]
\end{equation}
where $s = {R_0\over R} {G_A^{(0)}\over 2\sqrt3 G_A^{(8)}}$. Since $s$ is 
target independent, we can use the value measured for the proton to deduce
$\Gamma_1$ for any other hadron target simply from the flavour non-singlet
form factors, which obey relations from flavour $SU(3)$ symmetry.
From our calculations, eq.(9), we find $s\sim 0.66$ at $Q^2=10{\rm GeV}^2$,
while the central value of the SMC result (7) gives $s\sim 0.55$.

The form factors for a hadron $\BB$ are given by the matrix elements of the
flavour octet axial current. The $SU(3)$ properties are summarised by
\begin{eqnarray}
&&\langle \BB |J_{I I_3 Y}^{\bf (\rho)}|\BB \rangle =
\langle {\bf \rho}^\BB| J^{\bf (\rho)}|{\bf \rho}^\BB \rangle ~~ 
\langle I^\BB I_3^\BB ; I^{\bBB} I_3^{\bBB} | I I_3\rangle \cr
&&~~~\times \left(\matrix{{\bf \rho}^\BB &{} &{\bf \rho}^{\bBB} &{} 
&\Big| &{\bf \rho}
&{} \cr I^\BB &Y^\BB &I^{\bBB} &Y^{\bBB} &\Big| &I &Y \cr}\right) 
\end{eqnarray}
Here, ${\bf \rho}$ indicates the $SU(3)$ representation while $I, I_3$ and $Y$ 
are the isospin and hypercharge quantum numbers. 
The term $\langle {\bf \rho}^\BB| 
J^{\bf (\rho)}|{\bf \rho}^\BB\rangle$ is a reduced matrix element,
while the other factors are $SU(2)$ and $SU(3)$ 
Clebsch-Gordon coefficients.

If we now take the hadron $\BB$ to be in the {\bf 10} representation,
then since
\begin{equation}
{\bf 10} \times {\bf \bar{10}} = {\bf 1} + {\bf 8} + {\bf 27} + {\bf 64}
\end{equation}
the matrix element of the (octet) current contains just one reduced
matrix element. In contrast to the case of $\BB$ in the octet 
representation as for the proton or neutron, there is no $F/D$ ratio
involved (recall ${\bf 8} \times {\bf 8} 
= {\bf 1} + {\bf 8} + {\bf 8} + {\bf 10} + {\bf 27}$ gives two
reduced matrix elements). This means that the ratio of $\Gamma_1$
for decuplet states can be predicted as a simple group-theoretic
number, up to the dynamical suppression factor $s$.
 
For example, for the $\Delta^{++}$, the matrix element of the 
current $J_{\mu 5R}^3$ is
\begin{equation}
\langle \Delta^{++} | J_{\mu 5R}^3 |\Delta^{++}\rangle =
\sqrt{{3\over10}} \langle {\bf {10}}|J^{\bf 8}|{\bf {10}}\rangle
\end{equation}
evaluating the C-G coefficient. A similar result holds for the 
current $J_{\mu 5R}^8$, with the C-G coefficient being 
$\sqrt{1\over10}$. Combining this with the equivalent results 
for the $\Delta^-$, we find the
ratio of the first moment of the polarised structure functions $g_1$
for the $\Delta^{++}$ and $\Delta^-$ is
\begin{equation}
{\Gamma_1^{\Delta^{++}}\over \Gamma_1^{\Delta^-}} =
{\sqrt{3\over10} + \sqrt{1\over10}\sqrt{1\over3}(1+4s) \over
\sqrt{3\over10} - \sqrt{1\over10}\sqrt{1\over3}(1+4s)} =
{2s+2\over 2s-1}
\end{equation}
The OZI prediction, $s=1$, would therefore be
$\Gamma_1^{\Delta^{++}}/\Gamma_1^{\Delta^-} = 4$.
This can be simply obtained in the valence quark model with the
assumption $\Delta u (\Delta^{++}) = \Delta d(\Delta^-)$.
However, substituting a suppression factor of $s\sim 0.66$ gives 
a much larger ratio 
$\Gamma_1^{\Delta^{++}}/\Gamma_1^{\Delta^-} \sim 10$,
while a value of $s$ nearer 0.5 would give an even bigger value.

We would therefore expect to find a quite spectacular deviation from 
the quark model expectation for this ratio of structure function 
moments. We can also show that the same result is obtained for the 
ratio $\Gamma_1^{\Sigma_c^{++}}/\Gamma_1^{\Sigma_c^{0}}$ for the 
charmed baryons $\Sigma_c^{++} = uuc$ and $\Sigma_c^{0} = ddc$. 
Of course, these examples have been
specially selected (because of the $2s-1$ factor) to show a particularly
striking difference from the simple quark model predictions.
However, as we now see, they are also the examples which can be
dynamically isolated in semi-inclusive DIS.

\section{Semi-Inclusive DIS}

The semi-inclusive DIS reaction $\mu N \rightarrow \mu hX$,
where $h$ is the detected hadron (a pion or $D$ meson), 
has distinct contributions from the current and target fragmentation 
regions, which we assume can be clearly kinematically distinguished. 
We require $h$ to be in the target fragmentation region
and furthermore that the hadron energy fraction $z$
should be large. In this limit, the reaction is well described by the
diagram below, in which a meson $h$ carrying a large fraction of the
target nucleon energy is emitted with the exchange of a Reggeon
$\BB$ with well-defined $SU(3)$ quantum numbers. A large rapidity
gap is required between $h$ and the inclusive hadrons $X$.
With a polarised beam and target, this kinematics\cite{SV2} allows 
us to measure the structure function $g_1^{\BB}$ of the exchanged
Regge trajectory $\BB$.
\vskip0.7cm
\vbox{
\centerline{
\psfig{figure=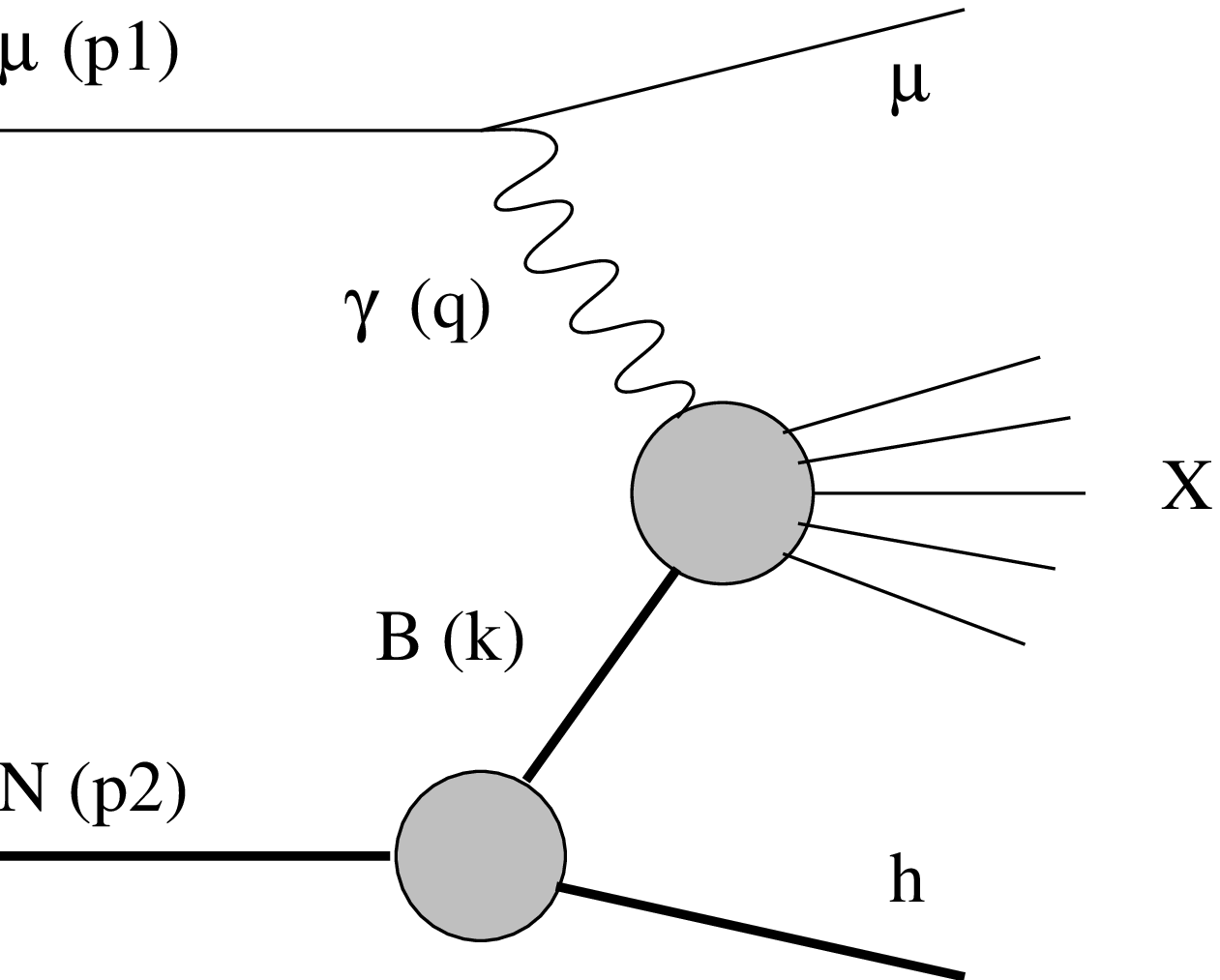,width=2in}}
}
\vskip0.7cm

Since the results of section 4 depend solely on the $SU(3)$ 
properties of the baryon $\BB$, they will still hold here despite
the fact that $\BB$ is interpreted as a Reggeon. If the target
is a nucleon $N$ and the detected hadron is an octet meson ($\pi$),
$SU(3)$ symmetry shows that $\BB$ belongs to a representation on 
the rhs of
\begin{equation}
{\bf 8} \times {\bf 8} = {\bf 1} + {\bf 8} + {\bf 8} + {\bf 10}
+ {\bf \bar{10}} +{\bf 27}
\end{equation}
Since the ${\bf 27}$ requires a 5-quark state, it is a good dynamical
approximation that the ${\bf 10}$ dominates the ${\bf 27}$.
However, there is no such argument for ${\bf 8}$ dominance
over the ${\bf 10}$. To isolate a unique representation
for $\BB$, we must therefore choose a combination of $N$ and $h$ 
giving $I_3,Y$ quantum numbers for $\BB$ which appear in the 
${\bf 10}$ but not in the ${\bf 8}$. This is 
satisfied by the $\Delta^{++}$ and $\Delta^-$, as in section 4.
The required ratio of first moments $\Gamma_1^{\Delta^{++}}/
\Gamma_1^{\Delta^-} = {2s+2\over 2s-1}$, where now $\Delta^{++}$ 
and $\Delta^-$ are Reggeons, is therefore obtained by comparing the 
reactions $\mu p\rightarrow \mu \pi^- X$ and $\mu n \rightarrow
\mu \pi^+ X$.

These symmetry considerations are easily pictured\cite{SV2} by
drawing quark diagrams for the $Nh\BB$ vertex. We also find 
that the same numerical ratio for the moments arises (as in section 4)
in processes in which a $D$ meson is detected, comparing the
reactions $\mu p\rightarrow \mu D^- X$ and $\mu n\rightarrow 
\mu D^0 X$. 

The dynamics of these semi-inclusive processes is described in detail
in \cite{SV2}. The first moment of the polarised structure function
for the Reggeon $\BB$ is found from the polarisation asymmetry of the
differential cross section in the target fragmentation region, i.e.
\begin{eqnarray}
&&\int_0^{1-z} dx~ x~ {d\Delta\sigma^{target}\over dx dy dz} \cr
&&= {Y_P\over 2} {4\pi \alpha^2\over s} \int dt\Delta f(z,t)
\int_0^1 dx_{\BB}~ g_1^{\BB}(x_{\BB},t;Q^2)~~~~~~~
\end{eqnarray}
Here, $x = {Q^2\over2p_2.q}$, $x_{\BB} = {Q^2\over2k.q}$, 
$1-z = {x\over x_{\BB}}$, $y = {p_2.q\over p_2.p_1}$, $t=k^2$
and $Y_P = {1\over y}(2-y)$. If we now take the ratio of
the two reactions described above, the factorised Reggeon
emission factor $\Delta f(z,t)$ cancels out, leaving the ratio 
of structure function moments $\Gamma_1$ predicted in section 4 to be given
simply by the ratio of the cross section moments.

All this can be precisely formulated\cite{SV2} in the language of
{\it fracture functions}, which have recently been applied both
to unpolarised and polarised semi-inclusive DIS (see \cite{SV2}
for citations).
In particular, the `Reggeon structure function' can be 
given a precise meaning in terms of a sum over partons {\it i} of the
(polarised) fracture functions $\Delta M_i^{hN}(x,z;Q^2)$ in 
the $z\sim 1$ region.

A detailed description of the dynamics of these semi-inclusive
processes and some further predictions will be given 
elsewhere\cite{SV2}. However, this brief sketch should be sufficient 
to demonstrate that there are exciting possibilities of discovering
relations between structure function moments which differ dramatically 
from simple quark-parton expectations and which could test the 
target-independent singlet suppression mechanism responsible for
the `proton spin' effect.

\end{document}